\documentclass[preprint, secnumarabic, amssymb, amsmath,nobibnotes, aps, prl,nofootinbib]{revtex4-1}

\usepackage{graphicx}

\newcommand{\half}{\mbox{\small{$\frac{1}{2}$}}}
\newcommand{\fourth}{\mbox{\small{$\frac{1}{4}$}}}

\begin{document}
\title{Geometric Phase, Curvature, and the Monodromy Group}
\author{B. H. Lavenda}
\email{info@bernardhlavenda.com}
\homepage{www.bernardhlavenda.com}
\affiliation{Universit$\grave{a}$ degli Studi, Camerino 62032 (MC) Italy}
\begin{abstract}
The geometric phase requires the multivaluedness of solutions to Fuchsian second-order equations. The angle, or its complement, is given by half the area of a spherical triangle in the case of three singular points, or half the area of a lune in the case of two singular points. Both are fundamental regions where the automorphic function takes a value only once, and a linear-fractional transformation tessellates the plane in replicas of the fundamental region. The condition that the homologues of the poles, representing vertices, be angles places restrictions on quantum numbers which are no longer integers, for, otherwise, the phase factors would become unity. Restriction must be made to regular singular points for only then will solutions to the differential equation be rational functions so that the covering group will be cyclic and the covering space be a \lq spiral staircase\rq. Many of the equations of mathematical physics, with essential singularities, become Fuchsian differential equations, with regular singularities, at zero kinetic energy.  Examples of geometric phase include the  phasor, the Pancharatnam phase of beams of polarized light in different states, the Aharanov-Bohm phase, and angular momentum with centripetal \lq attraction\rq. In the latter example, the phase is one-half the area of the lune, which disappears when the pole at infinity becomes an essential singularity thereby recovering the Schr\"odinger equation. The behavior of an automorphic function at a limit point on the boundary is analogous to the confluence of two regular singularities in a linear second-order differential equation to produce an essential singularity at infinity. 
\end{abstract}
\maketitle
\section{Introduction}

Quantum mechanics goes to great lengths to ensure that the wavefunctions are singlevalued. This means discarding terms in the solution to the Schr\"odinger equation that either blow up at the origin or diverge at infinity. Solutions of second-order differential equations which are rational lead to multivaluedness, and great efforts were spent,  in the late nineteenth century, to uniformize  the solutions so as to render them singlevalued. However, multivaluedness is not a stigma, and will explain numerous phenomena from the interaction of polarized beams to the Aharonov-Bohm effect. In this paper we treat multivaluedness from the theory of automorphic functions.

If a vector is parallel-transported around a closed curve it may not necessarily return as the same vector it started as. The effect is known as holonomy, and has been attributed to positive, Gaussian curvature~\cite{ONeill}. Holonomy also occurs when we solve a Fuchsian differential equation as a power series and take the analytic continuation around a regular singular point. We will, in general, not get back the solution we started with but one that differs from it by a phase factor. 

We will show that geometric phase is a manifestation of periodicity with respect to a group of motions of the tessellations of a disc, or half-plane, by lunes or curvilinear triangles, depending on whether the Fuchsian differential equation has two or three regular singular points, respectively. Functions whose only singular points are rational functions will be solutions to a Fuchsian differential equation of two singular points while the solutions of one with three regular points will not reduce to elementary functions, but rather can be expressed as a beta integral. 

 Differential equations containing only regular singular points, like the hypergeometric equation, have very little to do with the equations of mathematical physics~\cite{Gray}. Although the latter equations have a regular singular point at the origin they  possess an essential singularity at infinity that prevents the solution from diverging at infinity. The regular singular point at the origin has  linearly independent solutions, which are powers of the radial coordinate whose exponents are determined by the roots of the indicial equation. Their quotient is an automorphic function, whose inverse is a periodic function, that  will undergo a linear-fractional transformation and tessellate the plane with lunes, or curvilinear triangles. Quantum mechanics eliminates one of the solutions  on the basis that it blows up at the origin. However, this depends on the roots of the indicial equation.

 Because of a finite value of the kinetic energy, the other singular point at infinity is an essential singularity. The solutions are exponential rising and decaying functions of the radial coordinate. In order that the wavefunction be finite and singlevalued, the rising solution is excluded. The essential singularity arises as a coalescence of two regular singular points, and is analogous to the behavior of an automorphic function in the immediate neighborhood of limit points of the group of motions which tessellate the half-plane or principal circle. Therefore, if we allow for the multivaluedness of the  Schr\"odinger equation, its solutions will behave like automorphic functions far from the limit points on the boundary when we consider the limit of zero kinetic energy. 

In the next three sections, through the discussion of the phasor angle, the Pancharatnam phase of polarized light beams, and the Aharonov-Bohm phase, we will show that geometric phase requires positive Gaussian curvature so that the ratio of the area of  a curvilinear triangle to its angular excess is constant. Periodicity with respect to a group of motions tessellate the half-plane, or disc, which are natural boundaries upon which reside essential singularities. Periodicity requires at least two regular singular points, and the elliptic motion is a rotation. Non-integral values of quantum numbers are required in order that the group not reduce to the identity, corresponding to the equivalence class of null paths. These do not represent particles, whose quantum numbers must be integers, but, rather, are to be associated with resonances.

We then discuss \lq centripetal attraction\rq, for which the angular momentum varies over a continuous range of non-positive, and non-integral values. The quotient of the solution to the differential equation will take on each value only once in the lune, which is the fundamental region. This forms a dichotomy with quantum mechanics, where the angular momenta are discrete and space is continuous. We conclude the paper by reconstructing  the original Schr\"odinger equation: for negative kinetic energy the essential singularity is an exponential function, while for positive kinetic energy it is a circular function. As long as the kinetic energy vanishes, the Schr\"odinger equation, even in the presence of a potential, can be reduced to a Fuchsian form with multiple space scales.

\section{Phasor and the Construction of an Essential Singularity}

The linear-fractional transform, 
\begin{equation}
w=\frac{az+b}{cz+d},\label{eq:Mobius}
\end{equation} guarantees that the fundamental region will have the same number of poles and zeros, where $a,b,c,d$ are constants such that $ad-bc=1$. The difference between the number of zeros, $n$, and the number of poles, $p$, is given by
\begin{equation}
\frac{1}{2\pi i}\oint_{\mathcal{C}}\frac{f^{\prime}(z)}{f(z)}dz=n-p,\label{eq:n-p}
\end{equation}
where the contour $\mathcal{C}$ encloses all zeros and poles. Setting $f(z)=w$, with $w$ given by \eqref{eq:Mobius} we find
\begin{equation}
\frac{1}{2\pi i}\oint_{\mathcal{C}}\left(\frac{1}{z+b/a}-\frac{1}{z+d/c}\right)dz=0.\label{eq:0}
\end{equation}
The multipole moment of order $m$ is given by
\begin{equation}
\frac{1}{2\pi i}\oint_{\mathcal{C}}z^m\frac{f^{\prime}(z)}{f(z)}dz.\label{eq:multi}
 \end{equation}
The multipole moments are the analogs of essential singularties~\cite{Daniels}. Since equation \eqref{eq:multi} vanishes for an automorphic function; there can be no concentration of \lq charges\rq, which are analogs of zeros and poles, so that \eqref{eq:0} expresses charge neutrality.

\begin{figure}[htbp]
	\centering
		\includegraphics[width=0.60\textwidth]{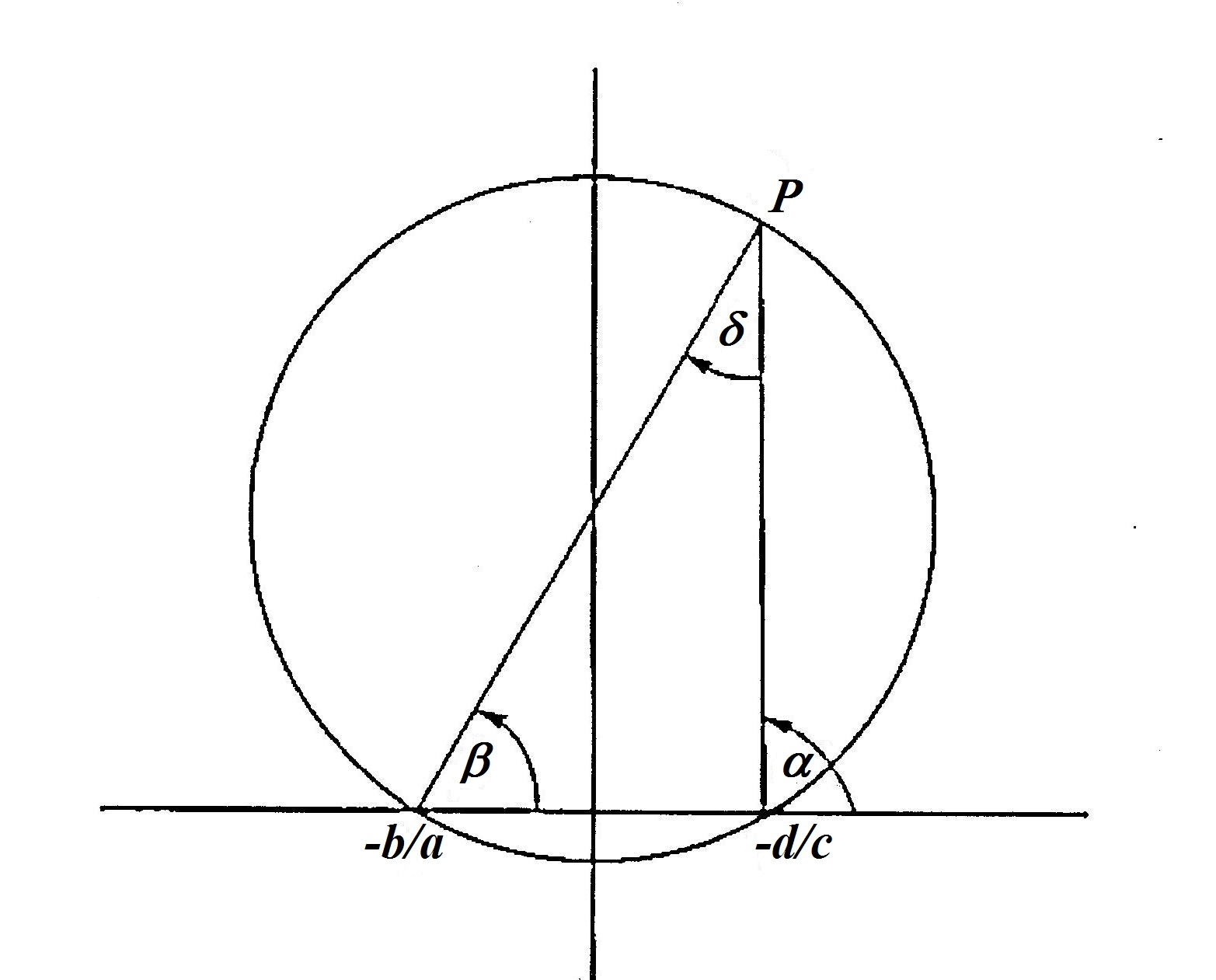}
	\caption{The contour is a circle passing through the pole at $-d/c$, and the zero $-b/a$.}
	\label{fig:contour}
\end{figure}

Real values of the coefficients in \eqref{eq:Mobius} will have the zero fall on the real axis. The contour in the $z$-plane for \eqref{eq:Mobius} is a circle passing through the pole at $-d/c$, and zero $-b/a$, as shown in Fig.~\ref{fig:contour}. The phase $\delta$ at point $P$ is the difference  between the angle $\beta$ and the exterior angle $\alpha$~\cite{Daniels},
\begin{equation}
\delta=\beta-\alpha. \label{eq:phasor}
\end{equation}
The lines of constant phase are circles which pass through $-b/a$ and $-d/c$. 

The  crucial, and new, point is to realize that by adding $\delta$ to both sides of \eqref{eq:phasor}, and adding and subtracting $\pi$ on the right-hand side give
\begin{equation}
2\delta=\delta+\beta+(\pi-\alpha)-\pi\ge0. \label{eq:anglex}
\end{equation}
The right-hand side is precisely the angle excess of a spherical triangle. We will soon appreciate that the phasor \eqref{eq:phasor} is the complementary angle to the Pancharatnam phase, \eqref{eq:Pan-bis}, to be discussed in the next section. 

The three angles of the triangle in Fig.~\ref{fig:contour}, $\delta=\lambda\pi$, $\beta=\mu\pi$, and $\pi-\alpha=\gamma\pi$, correspond to three \emph{regular\/} singular points, which by a linear-fractional transformation can be placed at $0$, $1$, and $\infty$. The simplest Fuchsian differential equation whose solutions do not reduce to elementary rational functions is one with three singular points. With $\beta$ at the origin, $\pi-\alpha$ at $1$, the phasor $\delta$ will be found at $\infty$. 

The automorphic function,
\begin{equation}
w=\int^{z}z^{\mu-1}(1-z)^{\gamma-1}\;dz, \label{eq:beta}
\end{equation}
is a beta integral, and satisfies the Fuchsian differential equation of second-order:
\begin{equation}
w^{\prime\prime}=\left(\frac{\mu-1}{z}+\frac{1-\gamma}{1-z}\right)w^{\prime}, \label{eq:Fuchs}
\end{equation}
where the prime stands for differentiation with respect to $z$. The value of the third angle, $\delta$ at $\infty$, can be  determined from the Schwarzian deterivative,
\[
\{w,z\}=\frac{1-\mu^2}{2z^2}+\frac{1-\gamma^2}{(1-z)^2}-\frac{2(1-\gamma)(1-\mu)}{2z(z-1)}. \]
Equating the numerator of the last term with the canonical form~\cite{Lehner},
\[\gamma^2+\mu^2-\lambda^2-1=-2(1-\gamma)(1-\mu),\]
 we find
\begin{equation}
\lambda=\pm(\gamma+\mu-1). \label{eq:pm}
\end{equation}
The negative sign will give the Euclidean result, 
\begin{equation}
\pi=\delta+\pi-\alpha+\beta, \label{eq:Euclid}
\end{equation}
which is the \emph{negative\/} of the phasor, \eqref{eq:phasor}, while the positive root in \eqref{eq:pm} will give the correct phasor, \eqref{eq:phasor}. This proves that the phasor belongs to spherical geometry, and not to Euclidean geometry.

\section{Pancharatnam's Phase for Polarized Light}

Berry~\cite{Berry} claims that Pancharatnam's phase~\cite{Pan} is one-half the solid angle subtended by a geodesic triangle on the Poincar\'e sphere. Without even knowing what the Pancharatnam phase is, it can safely be ruled out that the phase would be related to an interior solid angle when it is known that all deductions are made on the surface of the Poincar\'e sphere with absolutely no knowledge of the interior angles or points that the sphere encompasses~\cite{Shurcliff}. Moreover, any shape on the surface of the sphere that has the same area will have the same solid angle, and thus it need not be a geodesic triangle. In contrast, we will show that the complementary angle found by Pancharatnam is equal to half the area of a spherical triangle, given by the angle excess.

\begin{figure}[htbp]
	\centering
		\includegraphics[width=0.45\textwidth]{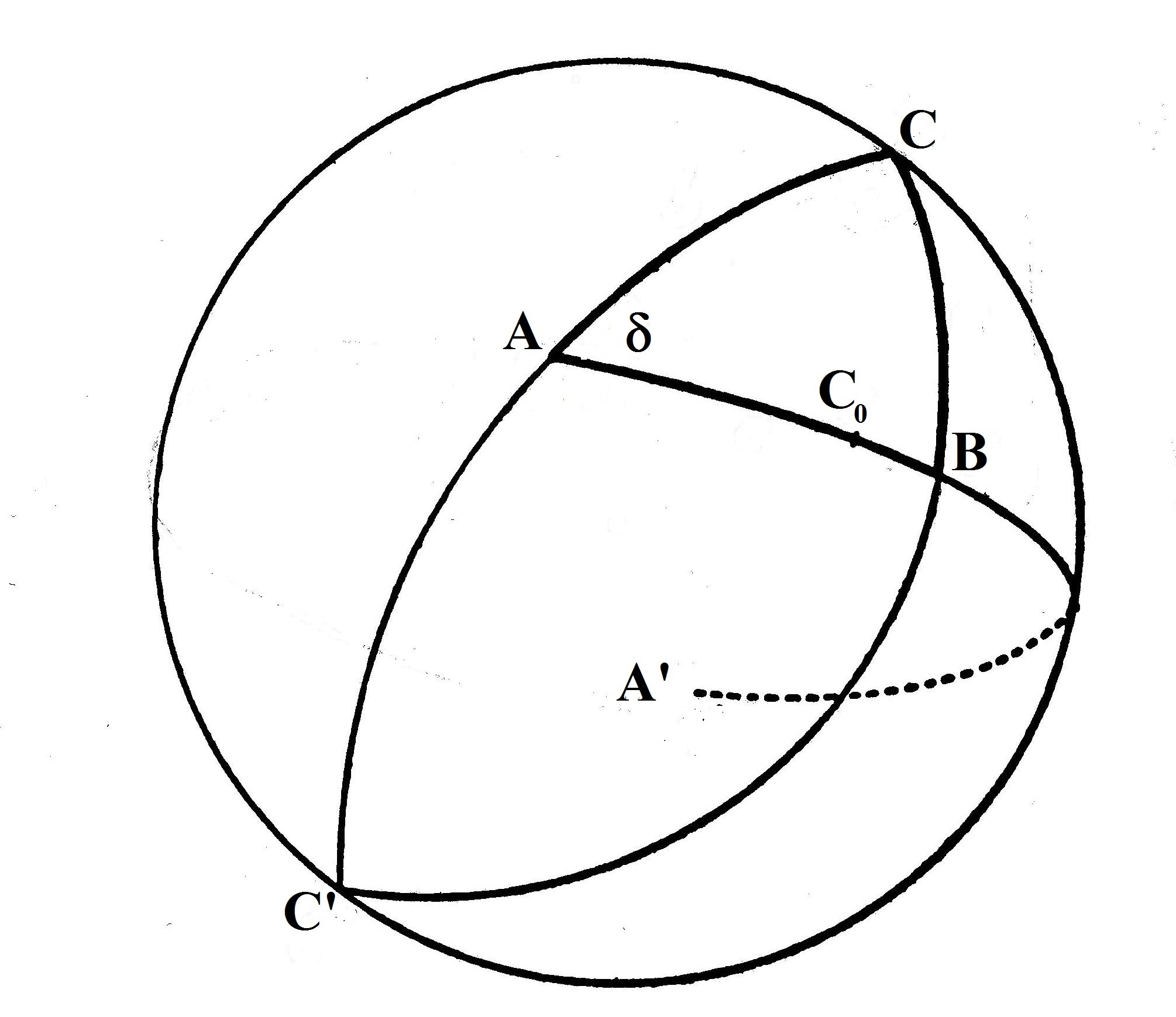}
	\caption{The phase $\angle C^{\prime}AB$ is determined by the angle excess of the triangle $\triangle BAC$ colunar to $\triangle C^{\prime}AB$. As $B\rightarrow C$ the the two beams will have opposite phases, while as $B\rightarrow C^{\prime}$, which is the opposite state of polarization to $C$, the phase difference will vanish.}
	\label{fig:Poincare}
\end{figure}

Pancharatnam considers a polarized beam $C$ to be separated into two beams in states of polarization $A$ and $B$, whose phase difference is the complementary angle to $\delta$. In reference to the phasor \eqref{eq:phasor}, $\delta$ will be equal to the difference in the internal angle $\angle ACB$ and the exterior angle $\angle ABC^{\prime}$,  
\begin{equation}
\delta=\angle ACB-\angle ABC^{\prime}, \label{eq:Pan}
\end{equation}
as shown in Fig.~\ref{fig:Poincare}. Expressing the exterior angle in terms of the interior angle, and adding $\delta=\angle BAC$ to both sides of \eqref{eq:Pan}, result in
\begin{equation}
2\delta=\angle BAC+\angle ACB+\angle ABC-\pi. \label{eq:Pan-bis}
\end{equation}
Equation \eqref{eq:Pan-bis} expresses twice the phase difference between the two beams in terms of the area of a spherical triangle given by its angle excess. 

Actually, Pancharatnam defines $\delta=\angle CAB$  as the phase difference which he expresses in terms of the triangle colunar to $\triangle ACB$, namely $\triangle AC^{\prime}B$. This is to say the angle, 
\begin{equation}
\angle C^{\prime}AB=\angle AC^{\prime}B-\angle ABC, \label{eq:delta}
\end{equation}
is the phasor, \eqref{eq:phasor}, being the difference between  the opposite internal angle and the external angle of the third angle of the spherical triangle. Adding the angle $\angle C^{\prime}AB$ to both sides of \eqref{eq:delta}, and adding and subtracting $\pi$ on the right-hand side yield:
\begin{equation}
2\angle C^{\prime}AB=\angle C^{\prime}AB+\angle AC^{\prime}B+\angle ABC^{\prime}-\pi, \label{eq:delta-bis}
\end{equation}
The right-hand side of \eqref{eq:delta-bis} is the area of the triangle $\triangle C^{\prime}AB$, and replacing the left-hand side by its complementary angle gives
\begin{equation}
\delta=\angle CAB=\pi-\half\left(\angle C^{\prime}AB+\angle AC^{\prime}B+\angle ABC^{\prime}-\pi\right), \label{eq:5.a}
\end{equation}
which is eqn (5.a) of Pancharatnam~\cite{Pan}.

As $B\rightarrow C$, the phase $\angle C^{\prime}AB\rightarrow\pi$,  and the beams will have opposite phases. This is analogous to the coalescence of the zero and pole to form a multipole. Alternatively, as $B\rightarrow C^{\prime}$, the opposite state of polarization to $C$, the beams in the state of polarization $A$ and $B$ will have zero phase difference.

Pancharatman then asks what happens when the split component $B$ tends to the opposite polarized state $A^{\prime}$ of the other polarized component $A$? As $B\rightarrow A^{\prime}$ and $\delta\rightarrow\Delta$, the latter will be given in terms of the area of the lune cut out by the great circles $AC_0A^{\prime}$ and $AC^{\prime}A^{\prime}$, which  is $2\angle C_0AC^{\prime}$. Hence,
\begin{equation}
\Delta=\pi-\angle C_0AC^{\prime}=\angle C_0AC, \label{eq:lune}
\end{equation}
is half the area of the lune formed from the great circles $AC_0A^{\prime}$ and $ACA^{\prime}$. When the area vanishes, the beams will have opposite phases, $\Delta=\pi$. Fig.~\ref{fig:Poincare} also illustrates Pancharatnam's observation that the emergent state of polarization $C$ can be obtained from the incident state of polarization $C_0$ when polarized light passes through a birefringent medium, which can be viewed as a rotation of the Poincar\'e sphere through an angle $\Delta$ in the counterclockwise direction about the $AA^{\prime}$ axis.

\section{The Aharonov-Bohm Effect}
The fringe shift in a field-free, but multivalued, region due to a non-vanishing vector potential was predicted by Ehrenberg and Siday~\cite{Siday}, and rediscovered by Aharonov and Bohm~\cite{AR} a decade later. Ehrenberg and Siday found it strange that an optical phenomenon would be caused by a flux, instead of a \emph{change\/} in the flux. Aharonov and Bohm insisted on the multivaluedness of the region in which the beams are travelling. 

Consider the Schr\"odinger equation with a vector potential, $\mathbf{A}$,
\begin{equation}
i\hbar\frac{\partial\psi}{\partial t}=\frac{1}{2m}\left(\mathbf{p}-\frac{e}{c}\mathbf{A}\right)^2\psi. \label{eq:Schrodinger}
\end{equation}
We want to see how close \eqref{eq:Schrodinger} comes to a Fuchsian equation. It becomes one when the phase transform,
\begin{equation}
\psi\longrightarrow e^{-(i/\hbar)Et}\psi, \label{eq:trans-1a}
\end{equation}
is introduced into \eqref{eq:Schrodinger} and  the Hamiltonian, $H$, is replaced by $H-E$, which does not \lq\lq produce a trivial, computable phase change in the solution of [\eqref{eq:Schrodinger}]\rq\rq~\cite{Simon}. The reason why it is not trivial is because the constant $E$ would bring in higher-order poles in the indicial equation and introduce an essential singularity into the Schr\"odinger equation [cf. eqn \eqref{eq:hydrogen-tris} below]. As we shall show in the last section, the elimination of $E$ is a  necessary condition to keep all singular points regular in the Schr\"odinger equation, \eqref{eq:Schrodinger}.

The radial Schr\"odinger equation then reduces to
\begin{equation}
\psi^{\prime\prime}+P\psi^{\prime}+Q\psi=0, \label{eq:Schrodinger-bis}
\end{equation}
where the prime denotes differentiation with respect to the radial coordinate, $r$, and
\begin{align}
P&=-2\frac{ie}{\hbar c}A \label{eq:P}\\
Q&=-\left(\frac{ie}{\hbar c}A^{\prime}+\frac{e^2}{\hbar^2c^2}A^2\right).\label{eq:Q}
\end{align}
With a change in the unknown $\psi\longrightarrow k\psi$, \eqref{eq:Schrodinger-bis} becomes
\begin{equation}
\psi^{\prime\prime}+\left(P+2\frac{k^{\prime}}{k}\right)\psi^{\prime}+\left(Q+P\frac{k^{\prime}}{k}+\frac{k^{\prime\prime}}{k}\right)\psi=0. \label{eq:Yoshida}
\end{equation}
	If $k$ satisfies \eqref{eq:Schrodinger-bis}, the coefficient of $\psi$ vanishes in \eqref{eq:Yoshida}. Rather, if the coefficient of $\psi^{\prime}$ vanishes, $P+2k^{\prime}/k=0$,  \eqref{eq:Yoshida} reduces to
	\begin{equation}
	\psi^{\prime\prime}+I\psi=0, \label{eq:normal}
	\end{equation}
	where
	\begin{equation}
	I=Q-\fourth P^2-\half P^{\prime}=Q+P\frac{k^{\prime}}{k}+\frac{k^{\prime\prime}}{k}, \label{eq:Schwarzian}
	\end{equation}
	is half  the Schwarzian derivative. Equation \eqref{eq:normal} is known as the normal form of the equation. 
	
	Equations with the same normal form are said to be equivalent, and $I$ is their invariant~\cite{Ince}. However, for the Schr\"odinger equation, \eqref{eq:Schrodinger-bis}, with coefficients \eqref{eq:P} and \eqref{eq:Q}, the invariant \eqref{eq:Schwarzian} vanishes identically. Therefore, \eqref{eq:Schrodinger-bis} is weakly equivalent to $\psi^{\prime\prime}=0$~\cite{Yoshida}, and there would be no invariant in the Aharonov-Bohm effect. Any function that has a vanishing Schwarzian derivative must be a linear-fractional transformation. And because a non-vanishing Schwarzian derivative is curvature~\cite{Tab}, we can conclude that \eqref{eq:Schrodinger} is not the correct equation to derive the Aharonov-Bohm effect~\cite{Wu}.
	
In fact, Aharonov and Bohm~\cite{AR} consider the wave equation outside the magnetic field region,
\begin{equation}
\left[\frac{\partial^2}{\partial r^2}+\frac{1}{r}\frac{\partial}{\partial r}+\frac{1}{r^2}\left(\frac{\partial}{\partial\vartheta}-i\alpha\right)^2+k^2\right]\psi=0, \label{eq:AR}
\end{equation}
where $\mathbf{k}$ is the wave vector of the incident particle, $\alpha=-e\phi/hc$, and $\phi$ is the total magnetic flux inside the circuit. By introducing the phase transformation
\[
\psi\longrightarrow e^{im\vartheta}\psi,\]
in \eqref{eq:AR} we can select a spherically symmetric solution by setting the magnetic quantum number $m=0$. Equation \eqref{eq:AR} then becomes the solution found by Tamm which is a Bessel function for $k^2>0$. According to Wu and Yang~\cite{Wu}, it has no meaningful solution if $k^2\le0$. However, it is precisely the equality that allows \eqref{eq:AR} to be transformed into the Fuchsian differential equation,
\begin{equation}
\psi^{\prime\prime}+\frac{1-(2\alpha)^2}{4r^2}\psi=0, \label{eq:AR-bis}
\end{equation}
provided $2\alpha<1$. According to Wu and Yang, the origin of this term is a monopole in the expression for the angular momentum,
\[\mathbf{L}=\mathbf{r}\times(\mathbf{p}-e\mathbf{A})-\frac{2\alpha\mathbf{r}}{r},\]
but the condition $\ell(\ell+1)\ge(2\alpha)^2$ would prevent the formation of a lune. We will now show their conclusion that \lq\lq the monopole does not possess strings of singularities in the field around it\rq\rq\ is inaccurate since analytic continuation about a regular singular point gives rise to a geometric phase.

Equation \eqref{eq:AR-bis} is valid about the singular point at the origin as well as the singular point at infinity. This can easily be shown by substituting $r=1/z$ in \eqref{eq:AR-bis} to get
\[\psi^{\prime\prime}+\frac{2}{z}\psi^{\prime}+\frac{1-(2\alpha)^2}{z^2}=0.\]
Then the substitution $\psi\rightarrow\psi/z$, will bring it  into the exact same form as \eqref{eq:AR-bis}. This shows that the fixed points at $r=0$ and $r=\infty$ are symmetrical.

\begin{figure}[htbp]
	\centering
		\includegraphics[width=0.50\textwidth]{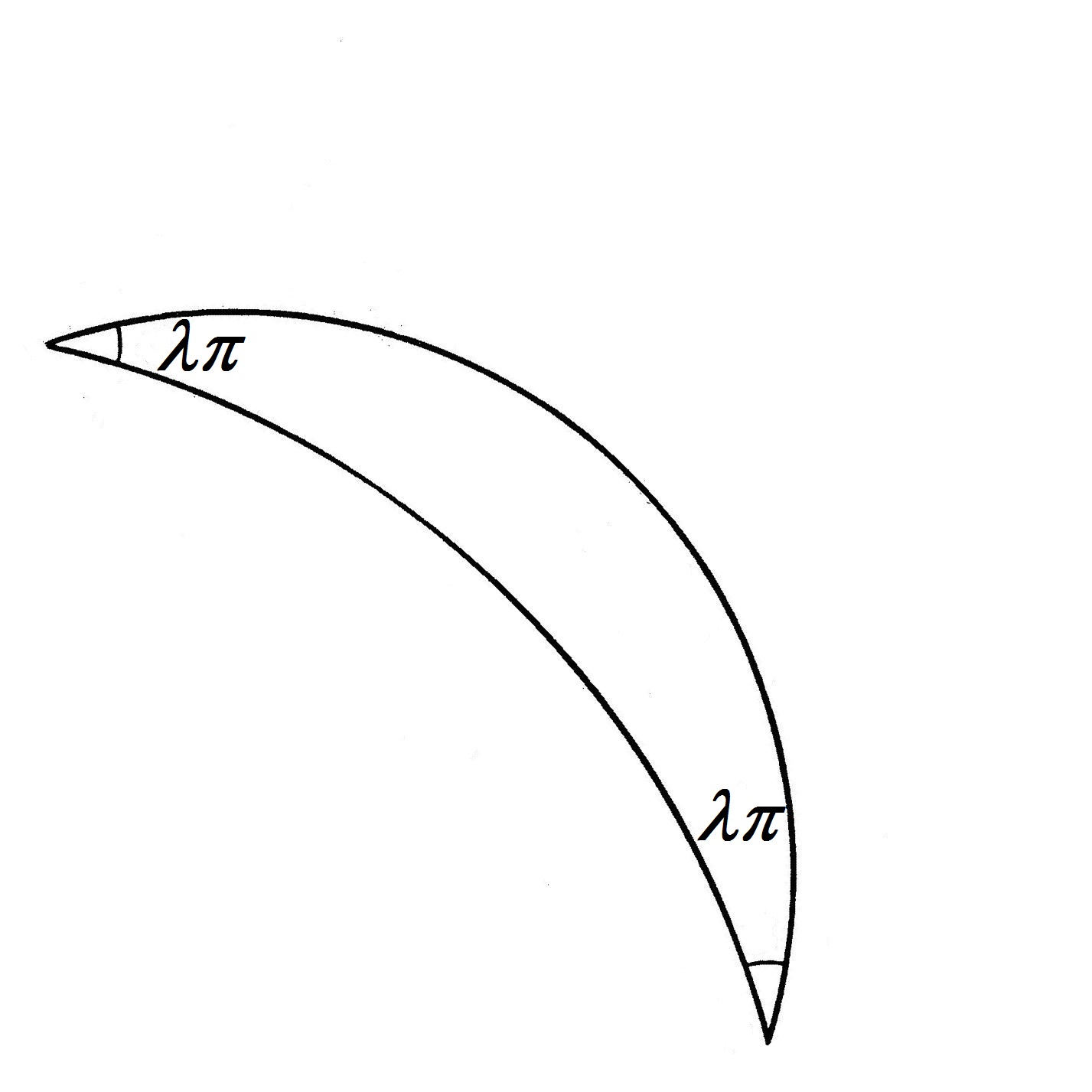}
	\caption{Two circular arcs intersect at an angle $2\alpha\pi$}
	\label{fig:lune}
\end{figure}

The two independent solutions to \eqref{eq:AR-bis} are:
\begin{equation}
\psi_1=r^{\half(1+2\alpha)}\hspace{60pt}\mbox{and}\hspace{60pt}\psi_2=r^{\half(1-2\alpha)}. \label{eq:AR-soln}
\end{equation}
Since  \eqref{eq:AR-soln} is multivalued, one solution would have to be rejected to preserve the singlevaluedness of the Schr\"odinger wavefunction. The quotient of the two solutions, \eqref{eq:AR-soln}, will undergo a linear-fractional transformation since any two independent solutions are linear combinations of any other pair of solutions. Analytic continuation about the origin, or infinity, will not give back the solution we started with. So by solving \eqref{eq:AR-soln} we have found functions automorphic with respect to a group of rotations. The group tessellates the upper half-plane, or disc, by lunes, of the form shown in Fig.~\ref{fig:lune}, where $r=0$ and $r=\infty$ correspond to the angular points of the lune.

Two circular arcs that cut out the lune  intersect at an angle $2\alpha\pi$. The area of the lune is $4\pi\alpha$. In terms of the phasor, the phase angle would be half this area, while Panacharatnam gives the phase as the complementary angle. Since we want the phase to vanish with the magnetic flux intensity, we choose the former and get
\begin{equation}
\delta=2\pi\alpha=\frac{2\pi e\phi}{hc}=\frac{e\phi}{\hbar c}=\frac{e}{\hbar c}\oint\mathbf{A}\cdot d\mathbf{r}. \label{eq:AR-phase}
\end{equation}
The phase factor,
\begin{equation}
\psi=e^{2\pi i\alpha}, \label{eq:factor}
\end{equation}
is the change in the wave function during a circuit of the solenoid. Equation \eqref{eq:AR-phase} says that when $\phi$ is an odd multiple of a fluxon, $hc/2e$, the two beams (one  bypasses  the toroidal magnetic and  the other passes through its hole) should exhibit a (maximum) phase difference of $\pi$ (mod $2\pi)$, i.e.,
\[\frac{2\nu+1}{2}2\pi\equiv\pi \hspace{30pt}(\,\mod 2\pi)\hspace{30pt}\nu=0,\pm1,\pm2\ldots\] This is what is seen in the interferogram that results from combining the beam with a coherent reference beam that avoids the magnetic field~\cite{Bate}. It is seen that integral quantization of the phase eliminates the phase factor, \eqref{eq:factor}, altogether. 
	
Denote by $\lfloor\alpha^{-1}\rfloor$ Gauss' bracket, which indicates the largest integer not exceeding $\alpha^{-1}$. Then $\varepsilon=e^{2\pi i/\lfloor\alpha^{-1}\rfloor}$ is an elliptic generator with period $\lfloor\alpha^{-1}\rfloor$. In other words, there will be $\lfloor\alpha^{-1}\rfloor$ distinct branches, or $\lfloor\alpha^{-1}\rfloor$ \lq steps\rq\ in the \lq spiral staircase\rq. The different branches are $g_n=\varepsilon^{n}g_0$, where $n=0,1,2,\ldots,\lfloor\alpha^{-1}\rfloor-1$ are the winding numbers. Each step can be regarded as a covering space corresponding to a particular branch of the multivalued function. In particular, for destructive inference of the beams, $\lfloor\alpha^{-1}\rfloor=2$, so that there is a single branch, and the surface is simply connected.
	
\section{Attractive Angular Momentum}

Many of the equations of mathematical physics can be transformed into Fuchsian differential equations at vanishing kinetic energy. Consider the spherical Bessel equation,
\begin{equation}
\left(\frac{1}{r^2}\frac{d}{dr}r^2\frac{d}{dr}-\frac{\ell(\ell+1)}{r^2}+k^2\right)\psi=0, \label{eq:Bessel}
\end{equation}
which can be transformed into \eqref{eq:normal}
where
\begin{equation}
I=k^2-\frac{\ell(\ell+1)}{r^2}. \label{eq:I}
\end{equation}
The Bessel equation, \eqref{eq:Bessel}, has a regular singular point at $r=0$, and an essential singularity at $r=\infty$. This can be seen by making the substitution $z=1/r$, and noting that the coefficient of $\psi$ has higher-order poles at $z=0$ [cf. eqn \eqref{eq:confluent-tris} below]. 

The indicial equation at the regular singular point, $r=0$, has  two independent solutions:
\begin{equation}
\psi_1=r^{\ell+1}\hspace{60pt}\mbox{and}\hspace{60pt} \psi_2=r^{-\ell}. \label{eq:soln-Bessel}
\end{equation}
The second solution $\psi_2$ is ordinarily discarded on the basis that it blows up at the origin. This makes $\psi$ it singlevalued. The quotient of the two solutions, 
\begin{equation}
s=\psi_1/\psi_2=r^{\lambda}, \label{eq:quotient}
\end{equation}  has a multivalued nature, and is automorphic with respect to a group of rotations that will tessellate the half-plane, or disc, with lunes, if and only if $k^2=0$. There can be no constant terms appearing in  \eqref{eq:I}, or the Schwarzian derivative [cf. eqn \eqref{eq:Schwarz-Bessel} below].

When $k^2\neq0$, there will be an essential singularity at $r\rightarrow\infty$. We may study this singularity by making the substitution $z=1/r$, and as $z\rightarrow0$, \eqref{eq:Bessel} will reduce to
\begin{equation}
\psi^{\prime\prime}+\frac{2}{z}\psi+\frac{k^2}{z^4}\psi=0. \label{eq:Bessel-bis}
\end{equation}
The solution to \eqref{eq:Bessel-bis} gives an essential singularity,
\begin{equation}
\psi=\sin(k/z), \label{eq:essential-sin}
\end{equation}
at $z=0$ which consists of a pole of infinite order. It is the limit point of two sequences of zeros, one on the positive real, and the other on the negative real, axis~\cite{Daniels}. Since the integrand of \eqref{eq:n-p} is
\begin{equation}
\frac{f^{\prime}(z)}{f(z)}=-\frac{k}{z^2}\cot\frac{k}{z}=-\frac{1}{z}+\frac{k^2}{3z^3}+\frac{k^4}{45z^5}+\cdots, \label{eq:n-p-bis}
\end{equation}
and introducing it into \eqref{eq:multi} shows that it has a \lq charge\rq\ of $-1$, a vanishing dipole moment, a quadrupole moment of $k^2/3$, a hexadecapole moment of $k^4/45$, etc.

The automorphic function, $s$,  has the Schwarzian derivative,
\begin{equation}
\{s,r\}=\frac{1-\lambda^2}{2r^2}=2I, \label{eq:Schwarz-Bessel}
\end{equation}
only in case of vanishing kinetic energy, $k^2=0$, where $\lambda=2\ell+1$. As we have already shown, the indicial equations will then be identical about $r=0$ and $r=\infty$, thereby reducing the second singular point from an essential to a regular one. This is necessary insofar as the analytic continuation of the solution about the singular point will not give back the solution that we started with, but, the product of analytic continuations about two singular points will give back the original solution. In other words, the group of rotations needs, at least, two generators whose product is the identity. In the case of two singular points, the generators will be inverses of one another. This is Riemann's condition for the \lq\lq periodicity of the function\rq\rq~\cite{Gray}, and the group generated by these matrices is the \lq monodromy group\rq, a term coined by Jordan.

When the two poles are regular, a simply closed circuit in the counterclockwise direction about $r=0$, described by the monodromy matrix,
\begin{equation}
\mathbb{S}_0=\begin{pmatrix} e^{2\pi i\ell} &0\\ 
0 & e^{-2\pi i\ell} \end{pmatrix}, \label{eq:mono}
\end{equation}
must be accompanied by a counterclockwise circuit about the other singular point at $r=\infty$,
\begin{equation}\mathbb{S}_{\infty}=\begin{pmatrix} e^{-2\pi i\ell} &0 \\
0 & e^{2\pi i\ell} \end{pmatrix}, \label{eq:mono-bis}
\end{equation}
in order that Riemann's condition must be fulfilled: 
\begin{equation}
\mathbb{S}_0\mathbb{S}_\infty=\mathbb{I}, \label{eq:Riemann}
\end{equation} so that the motions form a group, the monodromy group. Periodicity results in a multivalued function only for non-integral values of $\ell$. Integral values would reduce the monodromy matrices, \eqref{eq:mono} and \eqref{eq:mono-bis}, to the identity matrix, and destroy the tessellations of the half-plane, or disc, by lunes. This is the condition for constructive interference, which is no longer possible when the singular point at infinity becomes an essential singularity. The presence of an essential singularity destroys the periodicity with respect to the group.

The existence of a lune formed from two circular arcs with angle $\lambda\pi$ implies that $\lambda\le1$, or, equivalently $\ell\in[-\half,0]$. The centripetal repulsion $\ell(\ell+1)$ has now become \lq centripetal attraction\rq, $\ell(\ell+1)<0$. 

The Bessel differential equation, \eqref{eq:Bessel}, thus becomes identical to the Aharonov-Bohm equation, \eqref{eq:AR-bis}. The automorphic function $s=\psi_1/\psi_2$ can be written more generally as
\begin{equation}
S=\frac{as+b}{cs+d}, \label{eq:Mobius-bis}
\end{equation}
which gives a conformal representation of the $S$-lune upon the $s$-half plane. Inside the lune, which is the fundamental region, the automorphic function will take on any value only once. Thus, the linear-fractional transformation, \eqref{eq:Mobius-bis}, will transform two circles cutting at angle, $\lambda\pi$, into any two others intersecting at the same angle. This result has been known since the time of Kirchhoff~\cite{K}.

Thus, space and angular momentum have switched roles: the former is discontinuous while the latter is continuous in the interval $\ell\in[-\half,0]$. The geometric phase is now half the area of the lune, $\delta=(2\ell+1)\pi$. For $\ell=-\half$ the regular and irregular solutions, \eqref{eq:soln-Bessel}, coalesce, and the phase vanishes. At the other extreme, $\ell=0$, and the phase, $\delta=\pi$, in which the area of the lune becomes the area of a hemisphere, and the Schwarzian derivative, \eqref{eq:Schwarz-Bessel}, vanishes. The differential equation \eqref{eq:Bessel} becomes weakly equivalent to $\psi^{\prime\prime}=0$ so that there is no invariant~\cite{Yoshida}, exactly as in the case of the Schr\"odinger equation \eqref{eq:Schrodinger}.

\section{Reconstruction of the Schr\"odinger Equation}

For Fuchsian automorphic functions, accumulation, or limit, points occur on the principal circle or real axis of the half-plane~\cite{Ford}. Not all points on the boundary need be limit points of the group. If the automorphic function is not a constant, each limit point of the group is an essential singularity of the function. The behavior of an automorphic function at a limit point is analogous to the behavior of the Schr\"odinger equation in the immediate neighborhood of the point at infinity. We first establish the form of the essential singularity in the case of  negative kinetic energy,~\footnote{For positive kinetic energy the essential singularity is given by \eqref{eq:essential-sin}.} and then show that the Schr\"odinger equation can be reduced to Fuchsian form even in the presence of a potential at infinity provided the kinetic energy vanishes.

Consider the radial Schr\"odinger equation for the \emph{bound\/} states of the hydrogen atom,
\begin{equation}
\psi^{\prime\prime}-\left[\frac{\ell(\ell+1)}{r^2}-\left(\frac{\gamma}{r}-\frac{1}{4}\right)\right]\psi=0, \label{eq:hydrogen}
\end{equation}
where  the parameter $\gamma=1/kr_B$, and $r_B$ is the Bohr radius. As $r\rightarrow0$, \eqref{eq:hydrogen}
becomes
\begin{equation}
\psi^{\prime\prime}-\frac{1-\lambda^2}{r^2}\psi=0, \label{eq:hydrogen-bis}
\end{equation}
which has two independent solutions, \eqref{eq:soln-Bessel}. 

As $r\rightarrow\infty$, \eqref{eq:hydrogen} reduces to
\begin{equation}
\psi^{\prime\prime}+\frac{2}{z}\psi^{\prime}-\frac{1}{4z^4}\psi=0, \label{eq:hydrogen-tris}
\end{equation}
when the transformation $r=1/z$ is made. The two independent solutions are
\begin{equation}
\psi_1=e^{-1/2z}\hspace{60pt}\mbox{and}\hspace{60pt}\psi_2=e^{1/2z}. \label{eq:soln-hydrogen}
\end{equation}
On the condition that  $\psi$ must remain bounded, as $r\rightarrow\infty$, or $z\rightarrow0$, the second solution in \eqref{eq:soln-hydrogen} is eliminated. The solution to \eqref{eq:hydrogen} is given as a product of the first solutions in \eqref{eq:soln-Bessel} and \eqref{eq:soln-hydrogen} multiplied by the associated Laguerre polynomials.

The transcendental function, 
\begin{equation}
f(z)=\psi_2/\psi_1=e^{1/z}, \label{eq:f}
\end{equation}
 has an essential singularity at $z=0$, corresponding to $r=\infty$. It can be considered as a limit of a rational function which is the ratio of a pole of order $n$ at $z=0$ and a zero of order $n$ at $z=-1/n$~\cite{Daniels}. The ratio,
\begin{equation}
\lim_{n\rightarrow\infty}\frac{(z+1/n)^n}{z^n}=\lim_{n\rightarrow\infty}(1+1/nz)^n=e^{1/z}, \label{eq:f-bis}
\end{equation}
has a finite limit coinciding with  a transcendental function.

This occurs on the principal circle, or the positive real axis of the half-plane.\footnote{Points at infinity can be transformed to the principal circle by the linear-fractional transformation,
\[U(z)=\frac{iz+1}{z+i}.\]} The essential singularity thus consists of the merger of a pole at infinite order at $z=0$ and a zero of infinite order at $r=0-$. Introducing \eqref{eq:f} into the multipole moment \eqref{eq:multi}, shows that the only non-vanishing moment is $m=1$ so that the essential singularity has a dipole moment of $-1$. This permits us to interpret poles and zero as opposite charges~\cite{Daniels}. 

If equation \eqref{eq:hydrogen-bis} has two singular points  $r=0$ and $r=\infty$ there are no limit points of the group of motions that separate the plane~\cite{Ford}. By transforming the singular point at infinity into an essential singularity, where an infinite number of poles will cluster,  we introduce a boundary, either a principal circle or real axis.  The transform involves introducing the kinetic energy which is represented by the last term in \eqref{eq:hydrogen}. The essential singularity has a dipole moment, which is related to a bound state, such as in the Schr\"odinger equation for the hydrogen atom, \eqref{eq:hydrogen}, in contrast to   an unbound state as in Bessel's equation, \eqref{eq:Bessel}, which has an infinite number of moments.

Let us look for a solution to \eqref{eq:hydrogen} of the Fuchsian type, $\psi(r)=r^{\ell+1}\varphi(r)$. Then $\varphi(r)$ will be the solution to
\begin{equation}
\varphi^{\prime\prime}+2\frac{(\ell+1)}{r}\varphi^{\prime}+\left(\frac{\gamma}{r}-\fourth\right)\varphi=0. \label{eq:confluent}
\end{equation}
Introducing the Euler operator, $D=rd/dr$~\cite{SK}, \eqref{eq:confluent} can be reduced to the Fuchsian form:
\begin{equation}
D(D+\lambda)\varphi=-r\left(\gamma-\fourth r\right)\varphi. \label{eq:confluent-bis}
\end{equation}
The resonances, or roots of the left-hand side of the equation, are $0$ and $-\lambda$. This conferms that for small $r$, the solution should behave as $r^{-\lambda}$ [cf. eqn \eqref{eq:quotient}]. The stable manifold is parameterized by $\gamma$, the coefficient of the attractive Coulombian potential.

Solving \eqref{eq:confluent-bis} recursively, we get the power expansion
\[
\varphi=r^{-\lambda}\left\{1+\frac{\gamma}{\lambda-1}r+\frac{1}{2(\lambda-2)}\left(\frac{\gamma^2}{\lambda-1}-\frac{1}{4}\right)r^2+\cdots\right\},\]
or in terms of our original wavefunction,
\begin{equation}
\psi=r^{-\ell}\left\{1+\frac{\gamma}{2\ell}r+\frac{1}{(2\ell-1)}\left(\frac{\gamma^2}{2\ell}-\frac{1}{4}\right)r^2+\cdots\right\}.\label{eq:psi-soln}
\end{equation}
The idea of such power series solution is the same as Frobenius's \lq trick\rq\ of considering logarithms as limiting cases of powers. Logarithmic solutions are admissible and occur when the roots of the indicial equation are equal. Equation \eqref{eq:psi-soln} shows that it is an analytic function which has a branch pole of order $-\ell$ at $r=0$.

When we apply the same procedure to the fixed point at infinity by setting $r=1/z$, we get
\begin{equation}
D(D-\lambda)\varphi=-\frac{1}{z}\left(\gamma-\frac{1}{4z}\right)\varphi, \label{eq:confluent-tris}
\end{equation}
which is not an equation of the Fuchsian type. At vanishing kinetic energy, \eqref{eq:confluent-tris} can be reduced to a Fuchsian type of differential equation by a transcendental change of variables,
\[R=e^{-1/z}.\]
Introducing two radial coordinates, $R_0=R$ and $R_1=R\ln R$~\cite{SK}, \eqref{eq:confluent-tris} can be brought into the form:
\begin{equation}
\mathcal{D}\left(\mathcal{D}+\lambda\right)\varphi=\gamma\frac{R_1}{R_0}\varphi, \label{eq:confluent-iv}
\end{equation}
where the two-space scale operator, $\mathcal{D}=R_1\partial/\partial R_0$.

 There is an analogy between the essential singularity at infinity of differential equations, like \eqref{eq:Bessel} and \eqref{eq:Schrodinger}, and the limit point point of a group,  which is also an essential singularity~\cite{Ford}. The essential singularities of the group are the essential singularities of the automorphic function. The limit points either lie along the real axis in the half-plane, or on the principal circle. When an autormorphic function is subjected to linear-fractional substitutions of the group, they will fill the half-plane or principal circle with fundamental regions that do not overlap and without lacunae. However, in the immediate vicinity of a limit point, the automorphic function assumes any number of different values. The fundamental regions tend to cluster in infinite number about points on the principal circle or on the real axis. Thus, \emph{the behavior of the automorphic function at a limit point on the boundary is analogous to the confluence of two poles in a differential equation to produce an essential singularity at infinity\/}.

\end{document}